\documentclass[letterpaper]{article} 
\usepackage{aaai2026}
\usepackage{times}  
\usepackage{helvet}  
\usepackage{courier}  
\usepackage[hyphens]{url}  
\usepackage{graphicx} 
\usepackage{natbib}  
\usepackage{caption} 
\usepackage{threeparttable}
\usepackage{booktabs}
\usepackage{algorithm}
\usepackage{algorithmic}

\usepackage{newfloat}
\usepackage{listings}
\DeclareCaptionStyle{ruled}{labelfont=normalfont,labelsep=colon,strut=off} 
\floatstyle{ruled}
\newfloat{listing}{tb}{lst}{}
\floatname{listing}{Listing}
\title{Red Light, Grey Zone: A Multi-Perspective Interactive Narrative for Autonomous Driving Ethics}
\author{
Mengyi Wei\textsuperscript{\rm 1},
Nianhua Liu\textsuperscript{\rm 1},
Chenyu Zuo\textsuperscript{\rm 2},
Liqiu Meng\textsuperscript{\rm 1}
}

\affiliations{
\textsuperscript{\rm 1}Technical University of Munich\\
\textsuperscript{\rm 2}University of Augsburg\\
mengyi.wei@tum.de, nianhua.liu@tum.de, chenyu.zuo@uni-a.de, liqiu.meng@tum.de
}

\nocopyright
\begin{document}

\maketitle

\begin{abstract}
Autonomous driving ethics is not only an expert concern, but also a public issue involving risk, responsibility, and governance. However, non-experts often struggle to interpret these issues in concrete incidents, especially when responsibility is distributed across multiple stakeholders. This paper investigates interactive narrative as a public-facing method for eliciting situated ethical reflection on autonomous driving. We present \emph{Red Light, Grey Zone}, a web-based, multi-perspective interactive narrative prototype inspired by a real-world autonomous-driving incident. The prototype invites participants to compare stakeholder perspectives, examine scene materials, and make responsibility judgments in the face of ethical ambiguity. We report an exploratory user study (\(N=12\)) examining how differently non-experts responded to the prototype. Our analysis focuses on three dimensions of reflection: ethical cognition, responsibility-focused critical thinking, and multi-perspective reasoning. Exploratory pre-post results showed the strongest self-reported shift in responsibility-focused critical thinking among participants who completed the intended stakeholder-comparison process, while ethical cognition and multi-perspective reasoning showed positive directional trends. Qualitative findings further show how participants reflected on safety and market trade-offs, responsibility ambiguity, transparency and privacy, and governance gaps. Participants also used stakeholder comparison to corroborate evidence and, in many cases, broaden responsibility judgments from single-actor blame toward more distributed interpretations of accountability. Overall, the study suggests that multi-perspective interactive narratives may support non-expert reflection on accountability, evidence, and governance in AI-enabled systems.
\end{abstract}


\section{Introduction}

As Artificial Intelligence (AI) becomes increasingly embedded in everyday life, ethical concerns have become more visible, more consequential, and more difficult to separate from questions of governance. These concerns do not typically appear as a single, general problem of “AI ethics.” Rather, they emerge in specific sociotechnical settings where responsibility, risk, and value trade-offs are distributed across multiple actors. Autonomous driving is a particularly significant case because AI-driven decisions are enacted in public space, involve the possibility of physical harm, and require ongoing negotiation among companies, users, bystanders, and regulators. In this field, ethical debates often center on the question of responsibility: companies prioritize deployment and commercial viability, users value convenience and liability protection, pedestrians seek safety and fairness, while regulators must balance innovation, public accountability, and acceptable risk. These objectives can conflict, and their conflicts rarely admit a single correct solution. Ethical governance in autonomous driving therefore requires attention to sociotechnical and normative questions beyond technical performance alone.

However, a practical challenge is that meaningful public engagement with autonomous driving ethics remains difficult. Much of the existing discussion is shaped either by expert discourse or by simplified moral-dilemma framings, leaving limited room for understanding how ordinary people make sense of responsibility, risk, and governance in more situated ways. In this paper, we use non-experts to refer to members of the public who do not have formal training in the technical, legal, or ethical dimensions of autonomous driving, such as passengers, pedestrians, family members, and other affected citizens. Traditional awareness-raising approaches, including policy documents and educational lectures, often rely on abstract concepts and generalized explanatory frameworks. While such materials may communicate principles, they do not necessarily help non-experts understand how stakeholder positions differ in concrete situations, how risks are distributed, or how ethical conflict unfolds across a chain of events \cite{wei2025mapping}. As a result, public-facing reflection on autonomous driving ethics often remains broad and generalized, limiting the formation of more specific judgments about responsibility, accountability, and governance. Therefore, supporting non-expert reflection on these issues is an important challenge for AI ethics research and public-facing AI governance \cite{kwak2022influence}.

At the same time, existing work on autonomous driving ethics has largely relied on surveys, vignette-based moral dilemmas, trust and acceptance models, or scenario-based evaluations \cite{zhang2024public,ng2024acceptance,beckers2022blame,zhai2024responsibility}. These approaches have generated important insights into public preferences, blame attribution, and adoption attitudes. However, they tend to examine ethical responses as isolated judgments, with less attention to how interpretations develop across multiple stakeholder perspectives and stages of a contested event \cite{zhang2024public,stilgoe2021publicdialogue}. They also often foreground a single decision point, such as a collision dilemma, while giving less attention to the broader sociotechnical chain through which responsibility is interpreted, negotiated, and assigned \cite{rhim2021deeper,kirchmair2023regulate}. As a result, we still know relatively little about how non-experts reflect on autonomous driving ethics when they are asked to engage with ambiguity, compare stakeholder viewpoints, and reason about accountability across an unfolding scenario.

This study investigates interactive narrative as a public-facing engagement format and as a research method. It examines how non-experts make sense of controversial responsibility in autonomous driving. Interactive narrative enables participants to engage with unfolding events through choices, perspective shifts, and consequence-based progression. By situating ethical issues in concrete scenarios, it makes stakeholder positions, conflicting values, and possible consequences more legible \cite{naul2020story,buijsspanjers2020narrative,belman2010designing,dasilva2025ethicalreflection}. In autonomous driving, ethical disagreement emerges across events involving various actors and constraints. Interactive narrative presents ethical issues accessibly and enables closer examination of concerns that non-experts find important. It shows how interpretations of responsibility differ across stakeholder perspectives\cite{belman2010designing}. The study moves beyond asking if non-experts can identify ethical issues. It examines how multi-perspective engagement shapes non-expert reasoning about responsibility and governance in AI systems \cite{stilgoe2021publicdialogue,rhim2021deeper}.

Building on this rationale, we investigate how a \emph{web-based multi-perspective interactive narrative prototype} can support non-experts’ reflection on ethics in autonomous driving incidents:

\begin{itemize}
\item {\textbf{RQ1:}} What ethical themes become salient to non-experts when engaging with the interactive narrative about autonomous driving ethics?
\item {\textbf{RQ2:}} How does comparing stakeholder perspectives shape non-experts’ interpretations of ethical conflict in autonomous driving?
\end{itemize}

To address these questions, we develop Red Light, Grey Zone, an interactive narrative prototype grounded in a real-world autonomous driving ethics incident.\footnote{\url{https://incidentdatabase.ai/cite/8}}  The prototype is designed as a multi-perspective experience in which participants examine a contested autonomous vehicle incident through the perspectives of different stakeholders. Through a branching narrative, participants review scene materials, compare stakeholder accounts, and make judgments about responsibility and possible responses. By situating autonomous driving ethics within a specific sociotechnical context, the design allows non-experts to engage with ambiguity, conflicting interpretations, and distributed responsibility.

This paper makes three contributions. First, we introduce Red Light, Grey Zone, a web-based multi-perspective interactive narrative prototype that translates a real-world autonomous driving incident into situated scenarios, stakeholder accounts, and responsibility-judgment tasks. Second, we articulate a design approach for using interactive narrative to examine how non-experts reason about contested responsibility under incomplete information and divergent stakeholder accounts. Third, as part of the early-stage prototyping of AI ethics systems, through an exploratory user study (N=12), we provide empirical insight into what ethical themes become salient to non-experts and how stakeholder comparison shapes their interpretations of responsibility in an autonomous driving incident.

\section{Related Work}
\subsection{Interactive Narrative for Public Engagement with Autonomous Driving Ethics}
Research on autonomous driving ethics has revealed how public attitudes shape risk, blame, trust, and adoption. Most studies use surveys, vignettes, or models that isolate decisions or abstract trade-offs \cite{zhang2024public}. These approaches identify preferences and judgments but fail to capture how non-experts interpret ethical conflict as ongoing, socially situated events. This is critical, as ethical questions in autonomous driving arise across multiple stages—system design, deployment, accident interpretation, accountability, and governance\cite{stilgoe2021rejecting}. Abstract or static scenarios can be difficult for non-experts because they offer limited support for situating stakeholder positions, system behavior, and consequences within a coherent context.

To address these limitations, interactive narrative offers a useful response to the methodological challenge outlined above. It situates ethical issues within unfolding scenarios and enables participants to engage through choices, perspective shifts, and consequence-based progression. Prior work suggests that narrative-based and game-based formats can support engagement, reflection, and situated sense-making by making complex issues more concrete and easier to follow \cite{naul2020story}. Interactive narrative is therefore relevant as both a communication format and a research method for eliciting how participants interpret ambiguity, conflict, and responsibility in context. In this study, perspective-taking is treated as comparative reasoning rather than as empathy training.

Despite narrative and serious-game approaches being adopted in education and social studies, their use to study non-expert reasoning about autonomous driving ethics remains underexplored. Specifically, there is a limited understanding of how interactive narrative supports non-experts in expressing ethical concerns and responsibility judgments during a contested incident. This study directly addresses this gap, employing an interactive narrative prototype to reveal which ethical themes non-experts prioritize and how they make sense of responsibility in context. 

\subsection{Multi-Perspective Interaction in Autonomous Driving}
A core challenge in the ethics of autonomous driving lies in the fact that responsibility rarely rests with a single entity. Instead, it is dispersed among companies, engineers, users, pedestrians, investigators, regulators, and other affected members of the public. Consequently, the same incident may be interpreted in radically different ways depending on the perspective, whether viewed as a technical failure, a design choice, a governance issue, a communication breakdown, or an inequitable distribution of risk. In this sense, responsibility in autonomous driving is not only diffuse but also contentious\cite{stilgoe2021rejecting}. Prior work has shown that responsibility attribution in autonomous driving is shaped by contextual factors, including system capability, human control, accident framing, and public expectations. However, many existing studies still examine responsibility through single-shot judgments, such as blame allocation in a scenario, preference for a moral rule, or trust in a system’s decision. While such approaches are useful for identifying patterns in public evaluation, they offer less insight into how people may reinterpret responsibility when exposed to multiple stakeholder positions and conflicting accounts of the same event\cite{zhang2024public}.

For this reason, multi-perspective interaction functions as an analytic design choice in the present study. It allows us to examine how responsibility judgments are formed, challenged, and revised when the same incident is encountered from different stakeholder positions. This is particularly important in autonomous driving, where ethical disagreement often arises from divergent priorities, constraints, and interpretations \cite{rhim2021deeper}. A passenger may focus on safety and reliance on the system, a company on deployment and system limits, a regulator on public risk, and a family member on accountability and fairness. Comparing these positions can shift what participants notice, what they consider relevant, and how they assign responsibility. A multi-perspective interactive structure therefore provides a way to study contested responsibility in autonomous driving ethics. It supports closer examination of how non-experts compare stakeholder viewpoints, how ethical conflict becomes salient through contrast, and how responsibility interpretations shift during an unfolding incident. This motivates our use of multi-perspective interaction as a core design mechanism for examining non-experts’ ethical reflection and responsibility judgments \cite{dasilva2025ethicalreflection}.

\section{Design of the Multi-perspective Interactive Narrative}

\subsection{Overview and Design Goals}

Drawing on the literature reviewed above, we designed \textit{Red Light, Grey Zone} as a web-based, multi-perspective interactive narrative for examining how non-experts interpret ethical conflict in autonomous driving. The prototype was designed to provide a situated and interpretable representation of an ethically controversial incident. Our design specifically responds to two challenges identified in prior work: first, that non-experts often encounter autonomous driving ethics through abstract or fragmented explanations; and second, that responsibility in such incidents is distributed across multiple actors, making it difficult to understand from a single perspective.

The prototype is grounded in a real-world autonomous-driving incident in which an autonomous vehicle ran a red light, triggering public debate about safety governance and responsibility attribution. Public reports and incident documentation were used to preserve the core event structure and ethical tensions, including the red-light violation, disputed responsibility, and uncertainty about system behavior and public oversight. For interaction, these materials were adapted into fictionalized stakeholder accounts, dialogue, and scene materials. Contextual details, such as temporal ordering and location cues, were added to improve narrative coherence without changing the incident’s underlying responsibility structure. To avoid presenting a single authoritative frame, each stakeholder account was designed as partial and potentially biased, requiring participants to compare accounts before making a responsibility judgment. Additional details on scenario construction are provided in the supplementary materials.

A central design decision was to organize the prototype around four stakeholder perspectives: \textbf{company representative, eyewitness, company employee, and traffic authority}. These perspectives were selected because each captures a distinct standpoint from which the incident can be interpreted: organizational defense and risk management, immediate experiential observation, internal technical or institutional knowledge, and public regulatory accountability. The goal of this selection was to foreground contrasting positions that make responsibility attribution ethically and socially contested. In this way, the prototype uses stakeholder contrast not merely as a narrative device, but as a mechanism for examining how non-experts compare explanations, notice conflicting priorities, and revise their interpretations of responsibility.

This design directly supports the study’s two research goals. First, by engaging participants with a concrete yet contested incident, the prototype allows us to examine what ethical themes become salient to non-experts. Second, by presenting the incident through multiple stakeholder perspectives, it enables us to examine how participants’ interpretations of ethical conflict are shaped through comparison rather than through a single account.

\subsection{Design Rationale}

The prototype was designed to address two challenges identified in prior work. First, non-experts often encounter autonomous driving ethics through abstract explanations or simplified moral dilemmas, making it difficult to connect ethical concepts to concrete situations. Second, responsibility in autonomous driving incidents is distributed across stakeholders whose interests, institutional roles, and knowledge claims may conflict. To respond to these challenges, we designed the prototype as a situated, multi-perspective, and judgment-oriented interactive narrative.

More specifically, the design was guided by three principles. \textbf{P1. Situate ethical conflict in a concrete incident.} The prototype is organized around a single autonomous-driving incident adapted from a real-world case, allowing participants to consider how safety, accountability, and governance emerge through an unfolding situation. \textbf{P2. Highlight the perspectives of conflicting stakeholders.} The incident is presented from four perspectives: a company representative, an eyewitness, a company employee, and a traffic authority. These roles represent organizational defense, direct observation, internal knowledge, and public regulatory responsibility, making visible how the same event can be interpreted differently. \textbf{P3. Support responsibility reasoning under uncertainty.} Participants are positioned as investigators who examine scene materials, compare stakeholder accounts, and make a responsibility judgment under incomplete and sometimes conflicting information. Together, these principles operationalize ethical reflection as a sequence of actions: examining evidence, contrasting perspectives, recognizing uncertainty, and making a justified judgment.

\begin{figure*}[t]
    \centering
    \includegraphics[width=0.7\textwidth]{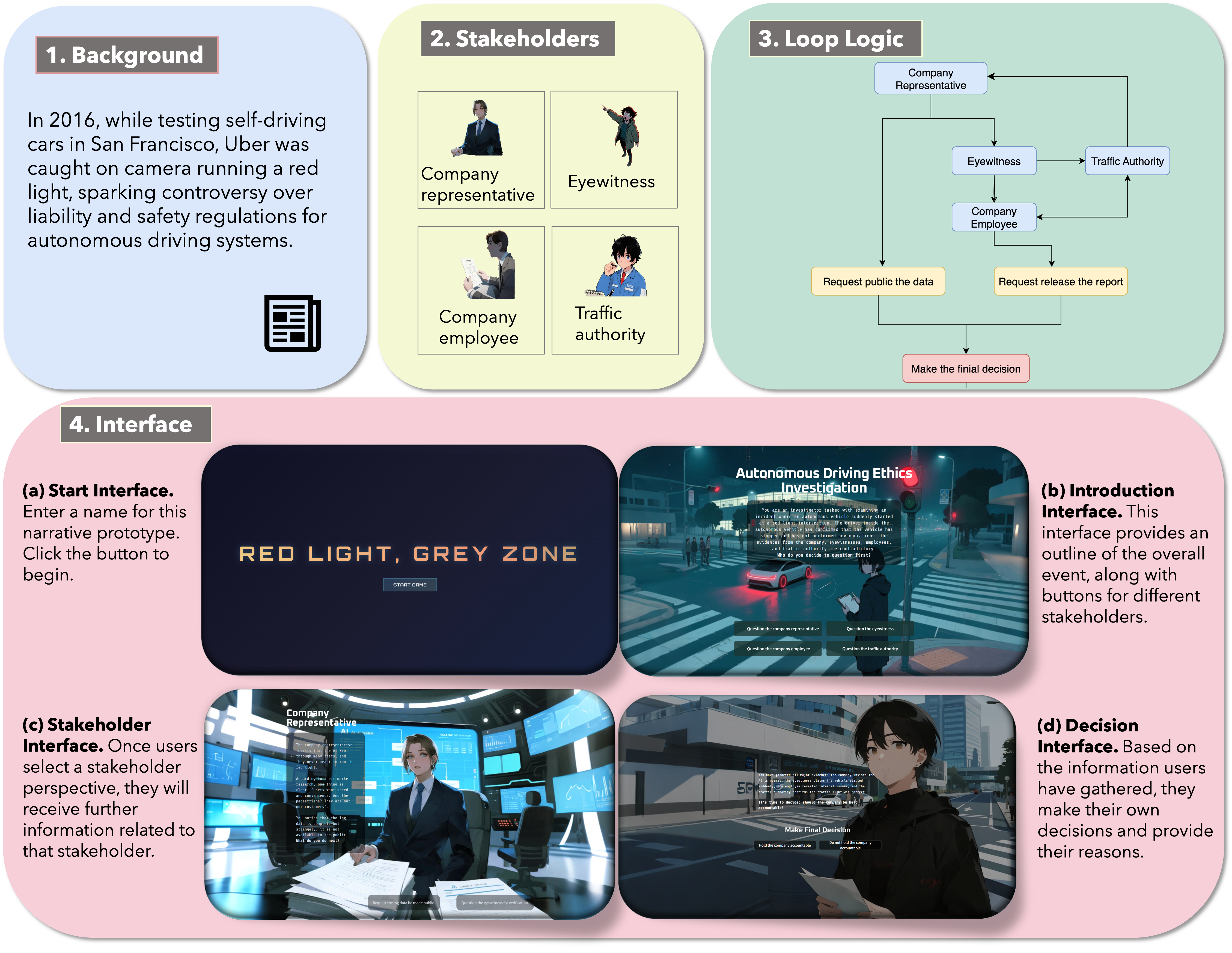}
    \caption{Overview of the design and implementation of \textit{Red Light, Grey Zone}. The figure shows four connected layers of the prototype: (1) the real-world incident background that grounds the narrative, (2) the four stakeholder perspectives used to structure contested interpretations of the event, (3) the interaction loop through which participants compare accounts and make a final responsibility judgment, and (4) the main interface stages that realize the web-based interactive narrative experience.}
    \label{fig:prototype-overview}
\end{figure*}

\subsection{Prototype Structure and Interaction Flow}

Figure~\ref{fig:prototype-overview} summarizes how the design rationale is translated into the prototype structure. The prototype consists of four connected components: an incident background, selected stakeholder perspectives, an interaction loop, and interface stages. Together, these components implement the three design principles described above.

First, the prototype is anchored in a real-world incident involving an autonomous vehicle running a red light (Figure~\ref{fig:prototype-overview}, top left). This incident provides a concrete entry point for reflecting on responsibility attribution, safety governance, and the public legitimacy of autonomous driving systems.

Second, the incident is presented through four stakeholder perspectives: \textit{company representative}, \textit{eyewitness}, \textit{company employee}, and \textit{traffic authority} (Figure~\ref{fig:prototype-overview}, top center). These perspectives were selected to foreground different interpretive positions rather than to provide an exhaustive map of all affected actors.

Third, participant interaction is organized as a loop in which users move across stakeholder accounts, request additional information, compare competing claims, and make a final judgment (Figure~\ref{fig:prototype-overview}, top right). Participants do not receive all information at once; instead, they work through partial and sometimes conflicting materials, which encourages them to assess credibility, notice uncertainty, and reconsider initial assumptions.

Fourth, these design choices are implemented through four interface stages: a start interface, an introduction interface, a stakeholder interface, and a decision interface (Figure~\ref{fig:prototype-overview}, bottom row). This staged progression guides participants from contextual orientation, to perspective comparison, and finally to explicit responsibility attribution. In this way, the prototype provides a structured setting in which non-experts can articulate ethical concerns and reconsider responsibility through multi-perspective engagement.

\subsection{Content Production and Implementation}
The prototype combines textual narrative with comic-style visual scenes to make the progression of the incident, the contrast between stakeholder accounts, and the points of ethical tension more immediately interpretable to non-expert participants. These visuals support comprehension of the unfolding scenario and highlight key moments of conflict, uncertainty, and investigation. Their function is illustrative and interpretive, focusing on how participants follow the incident and its ethical tensions. In this way, the visual layer complements the interactive narrative structure by making the incident easier to follow while maintaining a relatively low interpretive barrier.

The content production process was organized around the prototype’s incident-centered and multi-perspective design. For each scene, we first defined its narrative function in relation to the overall responsibility reasoning process, including what new information, uncertainty, or stakeholder position the scene was intended to introduce. We then developed scene-specific dialogue, materials, and participant options so that each perspective contributed a distinct interpretive standpoint rather than merely repeating previously presented facts. Comic-style visual panels were subsequently created to reinforce the main contextual or affective cues of each scene, such as conflict, doubt, institutional positioning, or evidentiary tension. To support iterative development and maintain stylistic consistency across scenes, we used generative AI tools to produce these illustrations. Their role was primarily practical, enabling rapid prototyping and visual coherence across the prototype. The study analyzed participants’ interaction with the narrative experience, rather than the generative AI production process itself.

As far as its implementation is concerned, the prototype was developed as a lightweight web-based application using HTML5, CSS, and JavaScript. This choice allowed the system to remain easily deployable and accessible through a standard browser, which was appropriate for an exploratory user study with non-expert participants. The narrative structure was implemented as a modular set of scene objects, each containing perspective-specific content, available actions, and branching transitions. Participant progression, selected options, and collected materials were managed through client-side state tracking, allowing the prototype to preserve the interaction flow across scenes. At each stage, the interface presents narrative text, a scene illustration, and two to four actionable options. These options determine the next step in the narrative and shape the information available to participants as they move toward a final responsibility judgment.
Overall, the prototype emphasizes situated ethical interpretation over technically detailed simulation. It provides a structured, accessible, and multi-perspective setting for non-experts to examine ethical conflict arising from a contested autonomous-driving incident.

\section{User Study}

We conducted an exploratory user study to clarify how non-experts engage with and interpret ethical issues in  \emph{Red Light, Grey Zone}, a web-based, multi-perspective interactive narrative about a contested autonomous-driving incident. The study aimed to determine what ethical topics became salient during interaction and how stakeholder perspective comparisons influenced views of responsibility. Instead of evaluating the prototype as a broad educational tool, we used it as a research instrument to elicit situated ethical reflection, responsibility reasoning, and governance-relevant trade-offs. All study procedures were approved by the IRB before recruitment.

\subsection{Participants}

We recruited 12 adult participants via university postings and internal channels. Participants were university-affiliated adults from mixed disciplinary backgrounds. None reported formal training or professional experience in autonomous driving, AI ethics, traffic law, or autonomous vehicle governance. Prior knowledge of autonomous driving was not required, allowing participants to engage as lay interpreters rather than domain specialists.

This convenience sample should not be considered statistically representative of the general public. However, it aligns with the exploratory objective of gaining preliminary insight into how non-experts interpret ethical conflict and responsibility through a multi-perspective interactive narrative. All participants signed informed consent forms prior to participation.

\subsection{Experimental Procedure}

We used a within-subjects pre-post design to capture participants’ self-reported shifts before and after interacting with the prototype. Because the study was exploratory and the sample size was small, pre-post comparison identified directional changes rather than strong causal or population-level claims. Sessions lasted about 30 minutes and consisted of three phases: a pre-interaction survey, the interaction session, and a post-interaction survey, with the full questionnaire provided in the supplementary materials.

\textbf{Pre-survey} (about 5 minutes). Participants rated their general knowledge and understanding of autonomous driving ethics on a 7-point scale. Items covered their self-perceived knowledge, understanding of ethical challenges, ability to spot ethical issues, and whether they tended to consider multiple viewpoints about autonomous driving. These items were the baseline for comparison.

\textbf{Interaction session} (about 10 minutes). Participants used Red Light, Grey Zone, which showed the incident background, stakeholder views, evidence, and a task for judging responsibility. They worked at their own pace and were told there was no single right answer. There was no time limit, but it usually took about 10 minutes.

\textbf{Post-survey} (about 15 minutes). After finishing the prototype, participants answered a post-survey with 7-point scale questions and open-ended questions. The scale questions asked if the prototype helped them understand ethical issues, notice complexity, see conflicts between safety, efficiency, and fairness, consider different sides, and reflect on the event. Open-ended questions asked them to describe ethical issues they saw, explain their responsibility judgment, and reflect on how comparing views changed their interpretation.

\subsection{Measures and Data Collection}

The study collected both quantitative and qualitative data. Quantitative data consisted of pre- and post-interaction Likert-scale responses. Because there is no established standard scale to measure non-experts’ reflection on autonomous driving ethics through interactive narrative, we developed study-specific questionnaire items informed by our research questions and the prototype’s goals. These items are intended as exploratory indicators, not validated psychometric scales.

We organized the questionnaire items into three analytic dimensions: \textit{ethical cognition}, \textit{responsibility-focused critical thinking}, and \textit{multi-perspective reasoning} (Table~\ref{tab:question-categories}). \textbf{Ethical cognition} refers to participants’ awareness and understanding of ethical issues in autonomous driving, including responsibility attribution, safety trade-offs, fairness, transparency, and privacy. \textbf{Responsibility-focused critical thinking} refers to participants’ self-reported tendency to question simplified explanations, examine competing claims about responsibility, and recognize evidentiary uncertainty in autonomous-driving incidents. This dimension was measured with a single paired item and is therefore interpreted as an exploratory item-level indicator rather than as a validated measure of general critical thinking ability. \textbf{Multi-perspective reasoning} refers to participants’ tendency to consider how different stakeholders may interpret the same incident differently.

\begin{table*}[t]
\centering
\begin{threeparttable}
\caption{Study-specific questionnaire items organized by analytic dimension.}
\label{tab:question-categories}
\begin{tabular}{p{0.34\textwidth} p{0.34\textwidth} p{0.18\textwidth}}
\toprule
\textbf{Pre-interaction item} & \textbf{Post-interaction item} & \textbf{Analytic dimension} \\
\midrule

Q3. I am familiar with ethical issues related to autonomous driving. 
& Q1. The prototype helped me reflect on ethical issues related to autonomous driving. 
& Ethical cognition \\

Q4. I understand what autonomous driving ethics means. 
& Q2. I became more aware of the complexity of ethical issues in autonomous driving. 
& Ethical cognition \\

Q5. I understand the ethical challenges involved in autonomous driving. 
& Q4. I noticed conflicts among efficiency, safety, and fairness in autonomous driving. 
& Ethical cognition \\

Q6. I can identify potential ethical issues in autonomous driving applications. 
& Q5. I am more likely to critically examine competing claims about responsibility in autonomous driving incidents. 
& Resp.-focused CT \\

Q7. I tend to consider multiple viewpoints when thinking about autonomous driving events. 
& Q3. The prototype helped me consider how different stakeholders may interpret the same autonomous driving incident. 
& Multi-perspective reasoning \\

\bottomrule
\end{tabular}

\begin{tablenotes}
\footnotesize
\item \textit{\textbf{Note.} The items are study-specific exploratory indicators rather than validated scales. Pre- and post-items were paired by analytic dimension, with different wording. Resp.-focused CT = responsibility-focused critical thinking} 
\end{tablenotes}

\end{threeparttable}
\end{table*}

These three categories were selected to capture complementary aspects of ethical reflection. Ethical cognition captures whether participants recognized and understood ethical issues in general. Responsibility-focused critical thinking captures whether participants noticed ambiguity and trade-offs, rather than accepting just one version of the event. Multi-perspective reasoning checks if participants compared different views, which is important to the prototype and RQ2.

As shown in Table~\ref{tab:question-categories}, ethical cognition was represented by three paired pre-post items. Responsibility-focused critical thinking and multi-perspective reasoning were each represented by one paired item. Because the latter two dimensions were measured with single paired items, the quantitative analysis of these constructs was treated as indicative rather than definitive. We therefore relied on qualitative analysis to provide a more detailed account of participants’ reasoning, especially regarding responsibility interpretation and stakeholder comparison.

The qualitative data included participants’ open-ended responses after the interaction, including descriptions of ethical issues, explanations of responsibility, and reflections on stakeholder perspectives. These responses were used to answer the research questions by identifying ethical themes important to participants and by examining how they understood responsibility through comparing stakeholder accounts.

\subsection{Data Analysis}

We used a mixed-methods approach combining descriptive quantitative analysis with qualitative thematic analysis. Given the exploratory nature of the study, the analysis aimed to characterize self-reported shifts and written reflections after participants interacted with the prototype, rather than to provide confirmatory evidence of intervention effectiveness.

For the quantitative analysis, we grouped Likert-scale items into the three analytic dimensions shown in Table~\ref{tab:question-categories}. Ethical cognition was computed as each participant's mean score across three paired items before and after interaction. Responsibility-focused critical thinking and multi-perspective reasoning were each represented by one paired item and were therefore treated as exploratory item-level indicators. We used paired-samples \textit{t}-tests to describe within-participant pre-post changes \cite{mara2012paired} and report Cohen's within-subjects effect size \(d_z=\bar{d}/S_d\) \cite{goulet2018review}, where \(\bar{d}\) is the mean paired difference and \(S_d\) is the standard deviation of the paired differences.

Because RQ2 concerns stakeholder comparison, the quantitative pre-post analysis focused on the \(n=10\) participants, from the full sample of \(N=12\), who engaged with at least three stakeholder perspectives before making the final responsibility judgment. The two limited-engagement participants were retained in the qualitative analysis as boundary cases. This distinction allowed us to describe self-reported shifts among participants who experienced the prototype's intended comparison mechanism while still examining variation in how participants engaged with that mechanism.

For the qualitative analysis, we conducted thematic analysis of participants' open-ended responses \cite{willig2017sage}. The analysis focused on two questions aligned with the research questions: what ethical themes participants foregrounded, and how they described the role of stakeholder comparison in responsibility judgment. Two researchers independently reviewed the responses, generated initial codes related to ethical concerns, trade-offs, stakeholder interpretations, uncertainty, and responsibility attribution, and then discussed and consolidated these codes into higher-level themes. Participant quotes were lightly edited for grammar and clarity without changing their substantive meaning. The goal was not to produce a comprehensive theory of autonomous-driving ethics, but to characterize recurring patterns in how participants made sense of the incident through the prototype.

Because the quantitative measures were study-specific and the sample was small, qualitative responses were used to interpret participants' reasoning rather than to validate the questionnaire results. In this sense, the quantitative analysis provides a coarse overview of self-reported shifts, while the qualitative analysis offers a richer account of the ethical themes and responsibility interpretations elicited by the interactive narrative.

\section{Results}

We report findings from participants’ interactions with \textit{Red Light, Grey Zone}. We first present exploratory pre-post questionnaire results to provide an overview of self-reported shifts among participants who completed the prototype’s intended stakeholder-comparison process. We then use qualitative analysis to examine the two research questions: what ethical themes became salient to participants after engaging with the interactive narrative (RQ1), and how comparing stakeholder perspectives shaped their interpretations of ethical conflict (RQ2).

\subsection{Overview of Exploratory Self-reported Shifts}

Before turning to the qualitative findings, we first report exploratory pre-post questionnaire results to provide an overview of participants' self-reported shifts after interacting with the prototype. Because the central mechanism of \textit{Red Light, Grey Zone} is stakeholder comparison, the quantitative analysis focused on participants who completed the intended comparison process. We defined completion as engaging with at least three of the four stakeholder perspectives before making the final responsibility judgment. Ten of the twelve participants met this criterion. The remaining two participants engaged with only one or two perspectives; rather than treating them as invalid cases, we retained them in the qualitative analysis as boundary cases that help characterize when and why multi-perspective engagement may be limited.

Table~\ref{tab:paired-results} reports exploratory paired-samples comparisons for the perspective-comparison subgroup (\(n=10\), from the full study sample of \(N=12\)). The clearest descriptive self-reported shift was observed for the responsibility-focused critical thinking item (\(+1.50\), \(t(9)=3.55\), \(p=.009\), \(d_z=1.25\)), with seven of ten participants reporting higher post-interaction scores. Ethical cognition and multi-perspective reasoning also showed positive directional changes; however, given the small sample and study-specific measures, these results should be interpreted as descriptive trends rather than confirmatory evidence.

These quantitative trends provide context for the qualitative findings reported below, especially participants’ attention to conflicting accounts, evidentiary uncertainty, and responsibility attribution. The positive trends in ethical cognition and multi-perspective reasoning are further unpacked through the ethical themes participants foregrounded and the ways they compared stakeholder perspectives. The two limited-engagement cases are discussed as boundary cases, illustrating that the prototype's reflective potential depends partly on how participants take up the stakeholder-comparison mechanism.

\begin{table}[t]
\centering
\small
\caption{Exploratory paired-samples comparisons for the perspective-comparison subgroup (\(n=10\), full sample \(N=12\)).}
\label{tab:paired-results}
\setlength{\tabcolsep}{3.2pt}
\renewcommand{\arraystretch}{1.12}

\begin{tabular}{lcccccc}
\toprule
\textbf{Measure} & \textbf{\(n\)} & \textbf{Diff.} & \textbf{\(t(df)\)} & \textbf{\(p\)} & \textbf{\(d_z\)} & \textbf{Imp.} \\
\midrule
Ethical cognition & 10 & +1.42 & 2.57(9) & .037 & 0.91 & 6/10 \\
Resp.-focused CT & 10 & +1.50 & 3.55(9) & .009$^{*}$ & 1.25 & 7/10 \\
Multi-persp. reasoning & 10 & +1.13 & 2.05(9) & .080 & 0.72 & 5/10 \\
\bottomrule
\end{tabular}

\vspace{3pt}
\begin{minipage}{0.98\columnwidth}
\footnotesize
\textit{\textbf{Note.} Diff. = post-pre score. Resp.-focused CT = responsibility-focused critical thinking. Ethical cognition averages three paired items. Resp.-focused CT and multi-persp. reasoning are single-item exploratory indicators. Imp. = participants with higher post scores, using all valid paired cases as the denominator. \(d_z\) = Cohen's within-subjects effect size. \(\alpha_{\mathrm{adj}}=.0167\). $^{*}p<.0167$.}
\end{minipage}

\end{table}

\subsection{RQ1: Salient Ethical Themes After Interactive Narrative Engagement}

Participants’ open-ended responses showed four recurring ethical themes: safety versus market incentives, ambiguity in responsibility attribution, transparency and privacy, and legal or institutional governance gaps. Together, these themes suggest that participants interpreted the incident through broader questions of accountability, evidence, and governance in autonomous driving, extending beyond the immediate traffic violation.

\textbf{Safety versus market incentives.}
Participants frequently treated the red-light incident as a concern about whether autonomous driving systems are deployed before sufficient validation or oversight. Several argued that safety should outweigh efficiency, convenience, or commercial pressure. As P01 stated, \textit{``I am more inclined to prioritize safety over efficiency.''} This suggests that participants understood autonomous driving ethics as a trade-off between technical performance, safety governance, and organizational incentives.

\textbf{Responsibility ambiguity.}
Nearly all participants identified responsibility attribution as central. Rather than locating responsibility in one actor, they described it as distributed across companies, drivers or users, system designers, and regulators. For example, P02 noted that \textit{``the company, developer, and regulator all share responsibility.''} Some also recognized institutional constraints; P03 suggested that employees may hold important information but lack authority to make it public. These responses frame responsibility as contested and institutionally mediated.

\textbf{Transparency, evidence, and privacy.}
Many participants emphasized data access, public records, and independent evidence when stakeholder accounts conflict. As P08 put it, \textit{``system logs should be released for independent review.''} At the same time, several noted that transparency may conflict with privacy protection or create risks of data misuse. Participants therefore treated transparency as a governance trade-off involving accountability, evidence access, and privacy constraints.

\textbf{Legal and institutional governance gaps.}
Participants also connected the incident to certification, accident investigation, legal standards, and regulatory oversight. Several called for clearer rules or third-party investigation mechanisms to support fair responsibility attribution and public trust. P10, for instance, called for \textit{``clearer rules before testing on public roads.''} This theme shows how participants moved from the immediate incident toward broader governance concerns.

Overall, the interactive narrative connected the vehicle's traffic-signal violation to wider concerns about safety governance, contested responsibility, evidentiary transparency, and institutional accountability.

\subsection{RQ2: How Stakeholder Comparison Shaped Responsibility Interpretation}

Participants' responses indicate that stakeholder comparison shaped responsibility interpretation in three ways: it encouraged corroboration across accounts, made conflicting priorities more visible, and broadened some judgments from immediate fault to distributed accountability. It also revealed boundary cases in which participants minimized perspective comparison because of prior beliefs or evidentiary preferences.

\textbf{From single account to corroboration.}
Several participants reported that comparing stakeholder perspectives helped them avoid relying on a single account. As P02 noted, \textit{``I would not rely on only one side of the story.''} Participants evaluated claims from the company representative, eyewitness, company employee, and traffic authority by considering which accounts appeared subjective, self-protective, or more verifiable. Some treated eyewitness testimony as useful but incomplete, while others placed greater weight on institutional records or technical explanations. This suggests that stakeholder comparison supported evidence-oriented responsibility reasoning: participants considered not only what was said, but also who said it and under what constraints.

\textbf{Making stakeholder conflict visible.}
The multi-perspective structure made ethical conflict more visible by presenting the same incident through different institutional and experiential positions. Participants identified tensions between corporate self-protection and public accountability, rapid deployment and safety validation, and transparency and privacy. P05 described this tension as \textit{``The company wants to protect itself, but the public needs accountability.''} These conflicts became more salient through comparison, suggesting that the prototype helped reveal how ethical disagreement emerges from divergent priorities and responsibilities.

\textbf{From individual blame to distributed responsibility.}
Many participants moved beyond attributing responsibility to a single actor and instead described responsibility as distributed among the company, system developers, regulators, and broader governance structures. As P02 wrote, \textit{``Responsibility should be shared among the company, engineers, and regulators.''} This was especially visible when participants connected the incident to system validation, regulatory oversight, or organizational decision-making. In this sense, stakeholder comparison broadened the interpretation of responsibility from immediate fault to sociotechnical accountability.

\textbf{Boundary cases: limited perspective exploration.}
Two participants, P04 and P09, consulted only one or two stakeholder perspectives before reaching a judgment. We treat them as boundary cases rather than invalid responses. P04 attributed responsibility to the company after confirming that the car ran a red light, explaining that companies often release immature technologies for profit and that additional viewpoints were unlikely to change this view. P09 was skeptical of subjective accounts and preferred verifiable evidence, treating some stakeholder testimony as potentially biased and emphasizing the need for extensive simulation and on-road validation before deployment.

These cases complicate a simple positive account of multi-perspective interaction: stakeholder comparison supported responsibility reasoning for many participants, but its role depended on prior beliefs, trust in evidence, and willingness to treat stakeholder accounts as relevant to judgment.

\section{Discussion}

Our findings suggest that the value of a multi-perspective interactive narrative lies less in teaching predefined ethical principles and more in making responsibility, evidence, and institutional constraints available for non-expert reasoning. In \textit{Red Light, Grey Zone}, participants did not engage with autonomous driving ethics only as an abstract question of whether an autonomous vehicle behaved correctly. Instead, many interpreted the incident through concrete concerns such as safety validation, commercial incentives, evidence access, information disclosure, privacy constraints, and regulatory accountability. We discuss three implications: how non-experts reasoned about autonomous driving ethics through situated responsibility problems, how interactive narrative can function as an elicitation method for public-facing AI ethics research, and what the multi-perspective structure reveals about stakeholder comparison.

\subsection{From Ethical Awareness to Situated Responsibility Reasoning}

The ethical themes identified in the study suggest that non-experts can engage with autonomous-driving ethics in concrete, governance-relevant ways when ethical issues are situated in a specific incident. Participants’ reflections clustered around safety versus market incentives, responsibility ambiguity, transparency and privacy, and legal or institutional governance gaps. Rather than treating these themes as separate opinions, we interpret them as indications that participants often understood the incident as part of a broader sociotechnical responsibility structure. The red-light violation became a starting point for considering how autonomous driving systems are developed, deployed, explained, and governed.

This finding shifts the emphasis from general “ethics awareness” toward situated responsibility reasoning. Participants did not necessarily use formal ethical vocabulary, yet they reasoned about core AI ethics issues through practical questions: whether the company had adequately validated the system, whether evidence should be disclosed, whether privacy should limit transparency, and whether independent investigation or regulatory oversight was needed. These concerns map onto broader questions of accountability and governance, but they appeared in participants’ responses as concrete judgments about evidence, responsibility, and institutional procedure.

Thus, incident-based interactive narratives may help make governance questions more accessible by translating them into interpretive problems: What evidence is available? Who has access to it? Which stakeholder account should be trusted? Who has the authority to disclose information, halt deployment, or assign responsibility? This suggests that public-facing AI ethics engagement need not begin with abstract principles alone; it can also begin with situated questions about how responsibility is made visible, contested, and governed.

\subsection{Interactive Narrative as an Elicitation Method for Public AI Ethics}

The study also points to the methodological value of interactive narrative for AI ethics research. Existing public-facing approaches often rely on surveys, vignettes, or abstract explanations. These methods can identify preferences or attitudes, but they provide limited insight into how participants navigate ambiguity, weigh evidence, or reinterpret responsibility as a scenario unfolds. By contrast, \textit{Red Light, Grey Zone} elicited final responsibility judgments as well as participants’ explanations of how they compared stakeholder accounts, evaluated evidence, and reasoned about institutional constraints.

In this sense, interactive narrative should not be understood only as an educational or communication format. It can also serve as an elicitation method for studying situated ethical interpretation. Because participants encounter information sequentially and through stakeholder perspectives, researchers can observe how they take up, reject, or reweight different accounts. This is particularly useful for studying domains such as autonomous driving, where responsibility is contested, evidence may be incomplete, and ethical disagreement is distributed across technical, organizational, legal, and public actors.

This methodological contribution is distinct from claiming that the prototype directly improved ethical competence. Our quantitative findings provide exploratory evidence of self-reported shifts among participants who completed the intended perspective-comparison process, especially in responsibility-focused critical thinking. However, the qualitative findings are more central to our contribution: they show how participants articulated responsibility, uncertainty, and governance concerns after working through the narrative. The prototype, therefore, functioned less as a tool for measuring whether participants learned a predefined lesson, and more as a structured setting for eliciting how non-experts make sense of contested responsibility in an AI-enabled sociotechnical system.

\subsection{The Role and Limits of Multi-Perspective Comparison}

The multi-perspective structure shaped participants’ interpretations of responsibility by making visible how the same autonomous-driving incident could be framed from different positions, interests, and information constraints. For many participants, comparing company, eyewitness, employee, and traffic authority perspectives appeared to support a shift from single-point blame toward more relational interpretations of accountability. Responsibility was not confined to the vehicle’s immediate behavior or to one actor’s decision; participants also linked it to organizational deployment choices, evidentiary access, disclosure practices, and regulatory oversight.

This suggests that multi-perspective interaction can help turn uncertainty into an object of reasoning rather than simply a barrier to judgment. When participants encountered conflicting accounts, they often asked which forms of evidence were more reliable, which stakeholders had incentives to withhold or frame information, and what institutional mechanisms might be needed to resolve uncertainty. In this way, the prototype made contested responsibility visible as a problem of evidence, position, and governance.

At the same time, our findings caution against assuming that access to multiple perspectives automatically leads to multiperspective reasoning. The boundary cases show that participants entered the narrative with prior assumptions, evidentiary preferences, and trust judgments that shaped whether they considered additional perspectives useful. P04 relied on a strong prior framing that companies may release immature technologies for profit, while P09 privileged verifiable evidence over subjective stakeholder testimony. These cases suggest that multi-perspective systems should not merely provide different viewpoints; they may also need to prompt users to reflect on why they privilege some accounts over others.

This also reframes the role of empathy in the prototype. Although perspective shifts may encourage participants to consider stakeholder constraints, our data do not justify strong claims about immersion, embodied role-taking, or empathy generation. The prototype positioned participants primarily as investigators comparing stakeholder accounts, not as fully inhabiting each stakeholder’s role. Its contribution is therefore better understood as perspective comparison rather than empathy training. This form of comparison foregrounds constraints, incentives, and information asymmetries as conditions under which responsibility is interpreted.

\subsection{Implications for Public-facing AI Governance}

These exploratory findings suggest implications for public-facing AI governance. Public engagement with AI ethics is sometimes framed as a problem of communicating principles to lay audiences. Our results suggest a complementary framing: non-experts may be better supported when ethical principles are translated into situated questions about responsibility, evidence, and institutional procedure. In the context of autonomous driving, participants’ concerns often centered on concrete governance mechanisms, including independent audits, traceable system logs, standardized disclosure, third-party investigation, and regulatory oversight. These mechanisms helped participants connect an individual incident to broader questions of accountability and legitimacy.

This matters for AI ethics because non-expert reasoning about governance does not necessarily begin with expert terminology. Participants did not need to use terms such as ``sociotechnical systems'' or ``algorithmic accountability'' to reason about related issues. Instead, they expressed concerns about who should disclose evidence, who should be trusted, how responsibility should be distributed, and what institutional safeguards should be in place. This suggests that interactive narrative may provide one way to bridge everyday reasoning and AI governance concepts without requiring non-experts to adopt expert language at the outset.

At the same time, the study also indicates that public-facing tools should be designed with limits in mind. A short interactive narrative cannot resolve complex governance questions, replace expert investigation, or represent all affected stakeholders. It can, however, provide a structured entry point for examining how non-experts interpret contested responsibility. In this respect, the prototype complements more formal policy documents, expert reports, and technical audits by making the interpretive and normative dimensions of autonomous driving incidents more accessible for public discussion.

\subsection{Limitations and Future Work}

This study has several limitations. First, the sample was small and recruited through convenience sampling, so the findings should be interpreted as exploratory rather than generalizable. The quantitative analysis focused on participants who completed the intended stakeholder-comparison process, while two participants with limited engagement were retained as qualitative boundary cases.

Second, the study captured short-term responses immediately after interaction. Future work should use delayed follow-up or longitudinal designs to examine whether interactive narratives based on incidents can sustain engagement with AI ethics.

Third, the questionnaire items were study-specific exploratory indicators rather than validated psychometric scales. Future studies should develop more robust measures of ethical cognition, responsibility-focused critical thinking, and multi-perspective reasoning.

Finally, the prototype was based on one autonomous-driving incident and four selected stakeholder perspectives. Future work could examine additional incident types, include a broader range of stakeholders, and test how narrative complexity and evidence access shape non-experts’ responsibility reasoning.

Overall, our study suggests that multi-perspective interactive narrative can support non-expert reflection on contested AI responsibility under uncertainty, rather than forcing a correct ethical answer.

\section{Conclusion}

This paper introduced \textit{Red Light, Grey Zone}, a web-based, multi-perspective interactive narrative adapted from a real-world autonomous-driving incident. The study examined how non-experts reason about ethical ambiguity, stakeholder conflict, and responsibility attribution in a situated scenario. Findings suggest that the prototype foregrounded governance-relevant concerns, including safety versus market incentives, responsibility ambiguity, transparency and privacy, and institutional gaps. Exploratory pre–post results indicated the strongest self-reported shift in responsibility-focused critical thinking, while qualitative findings showed that participants compared accounts, evaluated evidence, and often shifted from single-actor blame toward distributed accountability. Overall, the study suggests that multi-perspective interactive narrative may support non-expert reflection on responsibility, evidence, and governance in AI-enabled systems.

\section{Generative AI Use Statement}

Generative AI tools were used only to support rapid prototyping of illustrative visual panels. All study design, data analysis, and manuscript text were authored and verified by the authors.

\bibliography{aaai2026}


\end{document}